\documentclass[12pt]{article}
\usepackage[a4paper,margin=1in]{geometry}
\usepackage{setspace}
\usepackage{amsmath,amssymb}
\usepackage{graphicx}
\usepackage{booktabs}
\usepackage{multirow}
\usepackage{makecell}
\usepackage{siunitx}

\usepackage{subfig} 
\usepackage{pgfplots}
\pgfplotsset{compat=1.18}

\usepackage[hidelinks]{hyperref}
\usepackage{enumitem}\setlist{nosep,leftmargin=1.4em}
\usepackage{titlesec}
\titleformat{\section}{\large\bfseries}{\thesection.}{0.5em}{}
\titleformat{\subsection}{\normalsize\bfseries}{\thesubsection}{0.5em}{}

\newcommand{\papertitle}{Clinic-Oriented Feasibility of a Sensor-Fused Wearable for Upper-Limb Function}

\newcommand{\corrauth}{Aueaphum Aueawattthanaphisut}
\newcommand{\corremail}{aueawatth.aue@gmail.com}

\begin{document}
\begin{center}
{\LARGE \textbf{\papertitle}}\\[4pt]
{\large Thanyanee Srichaisak$^{2}$, Arissa Ieochai$^{2}$, \\ and Aueaphum Aueawattthanaphisut$^{1*}$}\\[3pt]
\small
$^{1}$Sirindhorn International Institute of Technology, Thammasat University, Pathum Thani, Thailand\\
$^{2}$King’s College Bangkok, Bangkok, Thailand\\
\textbf{Corresponding author:} \corrauth\ (\texttt{\corremail})\\

\end{center}

\section*{Structured Abstract}
\textbf{Background:} Upper-limb weakness and tremor (4--12\,Hz) limit activities of daily living (ADL) and reduce adherence to home rehabilitation.\\
\textbf{Objective:} To assess technical feasibility and clinician-relevant signals of a sensor-fused wearable targeting the triceps brachii and extensor pollicis brevis.\\
\textbf{Methods:} A lightweight node integrates surface EMG (1\,kHz), IMU (100--200\,Hz), and flex/force sensors with on-device INT8 inference (Tiny 1D-CNN/Transformer) and a safety-bounded assist policy (angle/torque/jerk limits; stall/time-out). Healthy adults ($n=12$) performed three ADL-like tasks. \emph{Primary outcomes}: Tremor Index (TI), range of motion (ROM), repetitions (Reps\,min$^{-1}$). \emph{Secondary}: EMG median-frequency slope (fatigue trend), closed-loop latency, session completion, and device-related adverse events. Analyses used subject-level paired medians with BCa 95\% CIs; exact Wilcoxon $p$-values are reported in the Results.\\
\textbf{Results:} Assistance was associated with lower tremor prominence and improved task throughput: TI decreased by $-0.092$ (95\%\,CI $[-0.102,-0.079]$), ROM increased by $+12.65\%$ (95\%\,CI $[+8.43,+13.89]$), and Reps rose by $+2.99$\,min$^{-1}$ (95\%\,CI $[+2.61,+3.35]$). Median on-device latency was 8.7\,ms at a 100\,Hz loop rate; all sessions were completed with no device-related adverse events.\\
\textbf{Conclusions:} Multimodal sensing with low-latency, safety-bounded assistance produced improved movement quality (TI$\downarrow$) and throughput (ROM, Reps$\uparrow$) in a pilot technical-feasibility setting, supporting progression to IRB-approved patient studies.\\
\textbf{Trial registration:} Not applicable (pilot non-clinical).

\vspace{0.3cm}
\noindent\textbf{Keywords:} wearable assistance; surface EMG; IMU; tremor suppression; rehabilitation; feasibility

\section{Introduction}\label{sec:intro}

\subsection*{Clinical Background and Burden}
Upper-limb neuromuscular disorders (NMD) and myopathies encompass inflammatory (e.g., polymyositis, dermatomyositis), dystrophic (e.g., Duchenne/Becker), metabolic, mitochondrial, toxic, and neurogenic involvement. Across etiologies, patients commonly experience four recurring pain points that limit activities of daily living (ADL): \textbf{weakness}, \textbf{tremor} with prominence in the \mbox{4--12~Hz} band, \textbf{limited range of motion (ROM)}, and \textbf{fatigue}. Weakness and fatigability diminish task throughput and precision; tremor injects unintentional motion energy that destabilizes grasp and reach; limited ROM restricts arm positioning and compensatory strategies. These deficits reduce independence in feeding, hygiene, dressing, and instrumental ADL, and they often undermine adherence to home rehabilitation because tasks are uncomfortable, inefficient, or discouraging.

\begin{figure} [h]
    \centering
    \includegraphics[width=1\linewidth]{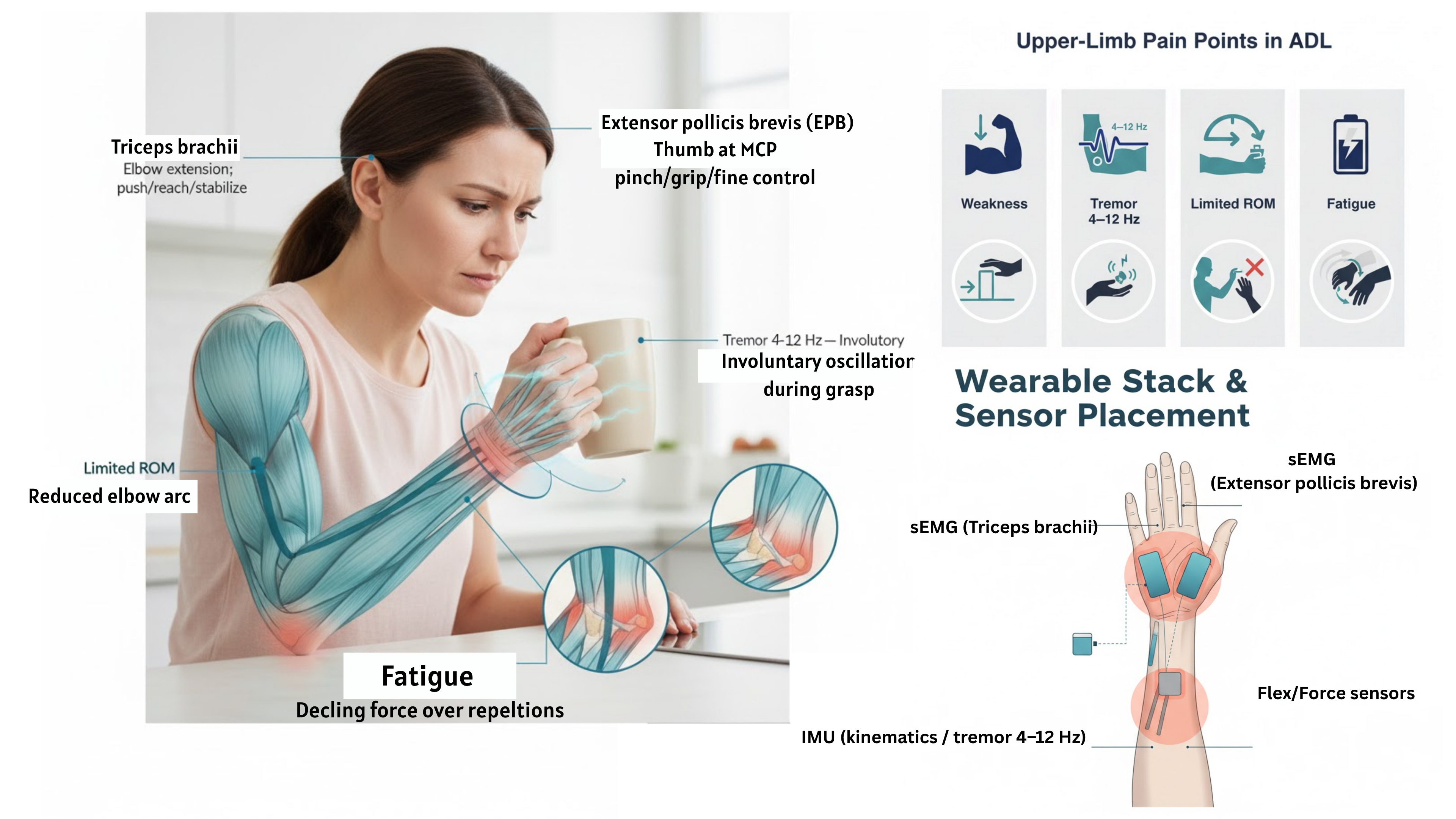}
    \caption{Pain points in upper-limb ADL during a grasp task: weakness, tremor (4–12 Hz), limited range of motion (ROM), and fatigue. The composite shows a realistic mug grasp with AR overlays—pain highlights at the lateral elbow and thumb MCP, tremor motion trails, a goniometric elbow arc indicating reduced ROM, and a fatigue icon—with zoom insets of tendon/joint stress.}
    \label{fig:painpoint}
\end{figure}

\subsection*{Pathophysiology Overview}
Muscle weakness in myopathy arises from fiber necrosis/regeneration cycles, fatty infiltration, connective tissue replacement, and disrupted excitation--contraction coupling. Inflammatory subtypes add immune-mediated fiber injury and pain; dystrophic subtypes reflect absent or dysfunctional cytoskeletal proteins (e.g., dystrophin/utrophin), altering force transmission. Neuromuscular junction and motor-unit remodeling reduce maximal voluntary contraction and change firing patterns; selective fiber-type loss (preferential type~II atrophy in many conditions) reduces high-power output. Central factors---including reduced descending drive, altered proprioception, and protective co-contraction against joint pain---further degrade smoothness and increase metabolic cost.

Tremor in NMD may be essential-like, postural/kinetic, or action-specific; in upper limb it typically concentrates around \mbox{4--12~Hz}. Pathways include cerebello-thalamo-cortical oscillations and peripheral reflex loops where mechanical resonance couples with stretch reflex gain. Clinically, the same band overlaps with hand-tool resonances and fine-manipulation frequencies, making tremor both noticeable and functionally disruptive during eating, drinking, writing, or smartphone use.

Peripheral and central fatigue coexist. \textit{Peripheral} mechanisms include impaired sarcolemmal excitability, metabolite accumulation, impaired calcium handling, and reduced cross-bridge cycling; \textit{central} mechanisms include reduced motor drive and altered afferent feedback. Surface EMG (sEMG) median frequency typically drifts downward with local muscular fatigue as conduction velocity decreases, providing a non-invasive marker to trend fatigue during tasks \cite{DeLuca1997JAB}.

\subsection*{Target Muscles and Biomechanical Relevance}
We focus on two muscles tightly coupled to common ADL tasks:

\begin{itemize}
  \item \textbf{Triceps brachii} (radial nerve): the primary elbow extensor; essential for \emph{pushing}, controlled lowering, reach stabilization, and support during load transfer. The long head also contributes to shoulder stability. Weakness yields reduced elbow extension torque, reliance on shoulder compensation, and difficulty generating controlled force against gravity or resistance.
  \item \textbf{Extensor pollicis brevis (EPB)} (posterior interosseous branch of the radial nerve): extends the thumb at the metacarpophalangeal (MCP) joint and assists retropulsion; it stabilizes the thumb for \emph{pinch}, \emph{grip}, and precision control. EPB dysfunction leads to collapse of the thumb column during grasp, impaired key pinch, and exacerbated kinematic jitter when tremor is present.
\end{itemize}

\noindent
Table~\ref{tab:anatomy} summarizes actions, innervation, functional roles, and typical deficits observed in clinic.

\begin{table}[t]
  \centering
  \caption{Target muscles: actions, innervation, ADL roles, and common deficits.}
  \label{tab:anatomy}
  \small
  \begin{tabular}{@{}p{2.9cm}p{2.3cm}p{3.0cm}p{4.3cm}@{}}
  \toprule
  \textbf{Muscle} & \textbf{Innervation} & \textbf{Primary action} & \textbf{ADL role / deficits}\\
  \midrule
  Triceps brachii & Radial n. & Elbow extension; joint stabilization & Push up from chair, reaching to shelf, controlling descent; weakness $\rightarrow$ poor extension torque, shoulder overcompensation, early fatigue\\
  EPB & PIN (radial) & Thumb extension at MCP; retropulsion & Pinch/grip, utensil control, smartphone tapping; deficit $\rightarrow$ thumb collapse, imprecise pinch, tremor-amplified jitter\\
  \bottomrule
  \end{tabular}
\end{table}

\subsection*{Clinical Phenotypes and Task Taxonomy}
\paragraph{Reaching and placing.} Requires coordinated shoulder flexion/scapular motion with elbow extension and forearm pronation/supination; triceps weakness manifests as undershoot or need for trunk lean.  
\paragraph{Pushing/transfer.} Chair rise and door push demand burst extension torque; fatigue leads to unsafe compensations.  
\paragraph{Grasp/pinch/fine manipulation.} Thumb MCP stabilization by EPB is critical; tremor in the 4--12~Hz band destabilizes precision tasks (spilling, mis-touch on screens).  
\paragraph{Hold \& reach-and-hold.} Isometric endurance is limited by peripheral fatigue and pain; micro-tremor rises with effort.

\subsection*{Assessment Landscape: Scales and Sensors}
Clinicians quantify impairment at multiple levels:

\begin{itemize}
  \item \textbf{Impairment level:} MRC grading for elbow extension and thumb extensors; handheld dynamometry for peak torque; ROM goniometry for elbow and thumb MCP; pain scales where relevant.
  \item \textbf{Activity level:} Fugl–Meyer upper-limb subtests, Nine-Hole Peg Test and Box-and-Blocks (dexterity), Action Research Arm Test, Disabilities of the Arm, Shoulder and Hand (DASH) questionnaire for patient-reported outcomes.
  \item \textbf{Physiology/sensor-derived:} sEMG amplitude features (RMS, MAV), median frequency slope for fatigue, and IMU-derived tremor prominence in \mbox{4--12~Hz}. Welch power spectral density (PSD) allows a ratio metric---the \emph{Tremor Index (TI)}---defined as the fraction of acceleration power in \mbox{4--12~Hz} over \mbox{0.5--20~Hz}, providing a unitless measure that scales across tasks and individuals. These signals can be acquired continuously during ADL-like tasks with a clinic-to-home wearable \cite{DeLuca1997JAB}.
\end{itemize}

\subsection*{Why the Triceps--EPB Pair Matters}
This pair anchors proximal \emph{force generation} (elbow extension torque) to distal \emph{precision control} (thumb stabilization). In practice, patients report that even when they can lift the forearm, they cannot \emph{use} the hand effectively: pushing a door is fatiguing, and grasping a cup or utensil is unstable. Proximal weakness also amplifies distal tremor because the limb operates closer to its mechanical limits, increasing reflex gains and reducing damping. Conversely, improved proximal stabilization often reduces distal oscillations perceived during precision tasks.

\begin{figure} [h]
    \centering
    \includegraphics[width=0.65\linewidth]{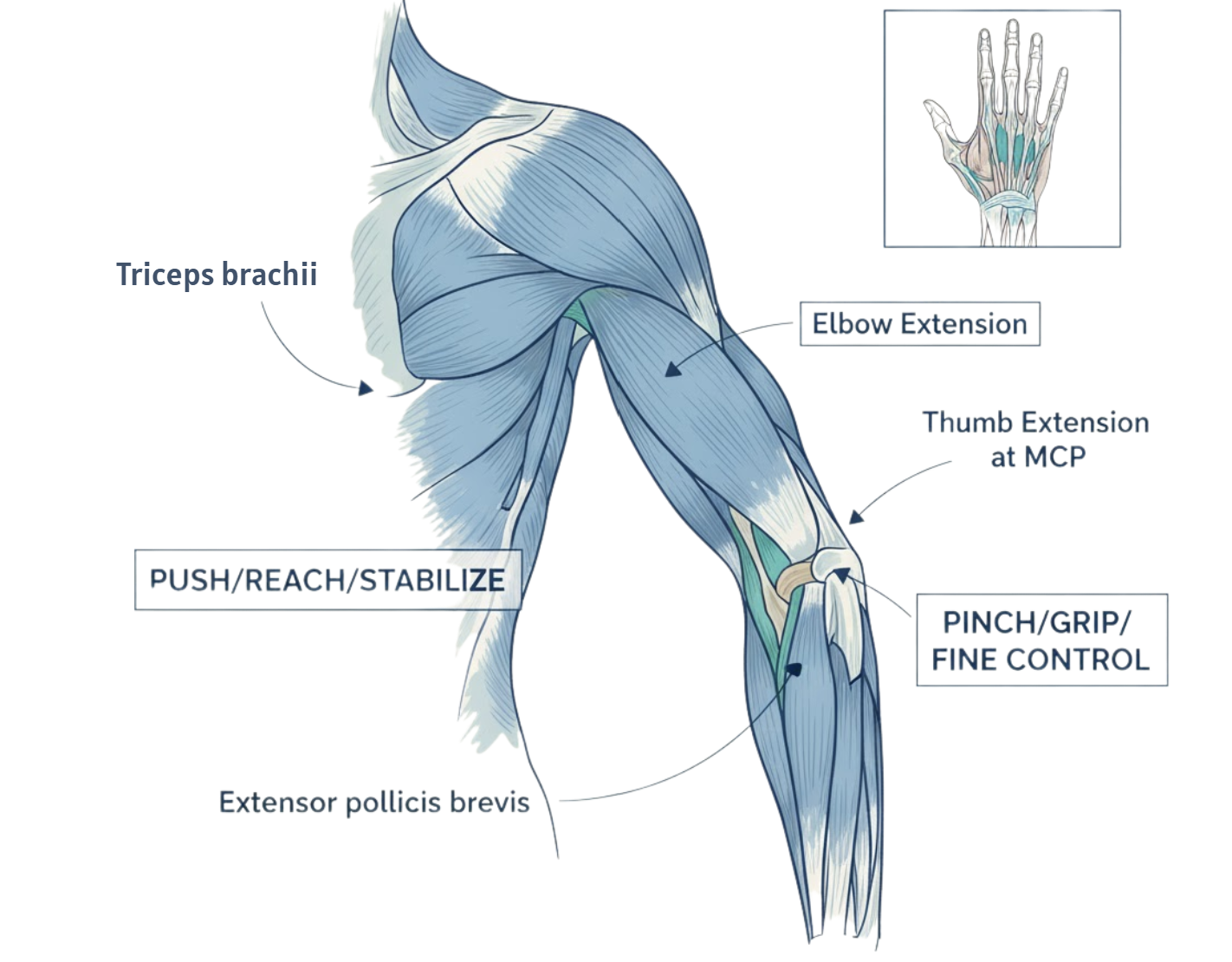}
    \caption{Target muscles and functional actions. \textit{Triceps brachii}: elbow extension (push/reach/stabilize). \textit{Extensor pollicis brevis}: thumb extension at the metacarpophalangeal (MCP) joint (pinch/grip/fine control). Anatomy is rendered with clean callouts to motivate the chosen outcomes (Tremor Index, ROM, repetitions).}
    \label{fig:anatomy}
\end{figure}

\subsection*{Medical Safety Considerations}
Any assistive strategy must respect joint integrity, soft-tissue tolerance, and pain. For the elbow, safe arcs and torque ramps avoid hyperextension or high jerk; for the thumb, MCP stress must remain within physiologic limits to prevent tendinopathy exacerbation. Fatigue management requires monitoring for declining force output and heat/discomfort under electrodes or straps. Closed-loop devices therefore benefit from \emph{safety envelopes}: angle limits, torque and jerk clamps, and stall/time-out logic that disengage assistance when abnormal loads or postures are detected.

\subsection*{Unmet Needs in Current Care}
Conventional therapist-guided exercise and home programs improve outcomes but hinge on adherence and feedback. Powered orthoses can amplify weak residual EMG but seldom fuse multimodal sensing for tremor quantification or provide continuous, low-latency assistance optimized for daily use \cite{Koo2014IROS}. Imaging and vision approaches (ultrasound, AR) aid diagnosis and monitoring yet are not intended for closed-loop, on-body assistance in naturalistic tasks \cite{Zhou2023TVCG,Ahmed2023Access}. Emerging embedded machine learning now enables on-device inference on low-power microcontrollers, but many systems lack transparent clinical endpoints, interpretable safety constraints, and pathways for clinic-to-home deployment.

\subsection*{Measurement Constructs Adopted in This Work}
To align engineering signals with clinical meaning, we adopt a compact outcome set:
\begin{enumerate}
  \item \textbf{Tremor Index (TI):} fraction of IMU acceleration power in \mbox{4--12~Hz} over \mbox{0.5--20~Hz}; lower is better (less tremor prominence).
  \item \textbf{Range of Motion (ROM):} elbow and thumb arcs (degrees) normalized to baseline; higher is better (greater excursion).
  \item \textbf{Repetitions (Reps min$^{-1}$):} cycles per minute for task-specific movements; higher is better (throughput).
  \item \textbf{EMG median-frequency slope:} trend of the median frequency over time; a less negative or stabilized slope suggests reduced fatigue accumulation \cite{DeLuca1997JAB}.
\end{enumerate}
These metrics can be computed in real time from lightweight sensors and summarized per session to support clinical decision-making.

\subsection*{Objective and Study Framing}
Given the burden and gaps above, we hypothesize that a \emph{sensor-fused} wearable focusing on triceps brachii and EPB can deliver \emph{low-latency, safety-bounded} assistance that improves movement quality (TI$\downarrow$) and throughput (ROM, Reps$\uparrow$) during ADL-like tasks, while monitoring fatigue trends. The present work is a \emph{pilot technical feasibility} evaluation in healthy adults designed to surface \emph{clinician-relevant signals} (TI, ROM, repetitions) and \emph{technical endpoints} (latency, completion, adverse events). Efficacy in patient populations and long-term adherence will be evaluated in subsequent IRB-approved trials. Details of the engineering pipeline (sensing, on-device modeling, safety control) and related literature are provided later and in prior engineering-focused work \cite{Auea2025NeurotremorENG}.

\section{Materials and Methodology}\label{sec:methods}

\subsection{Hardware Platform and Sensors}
\begin{figure} [h]
    \centering
    \includegraphics[width=1\linewidth]{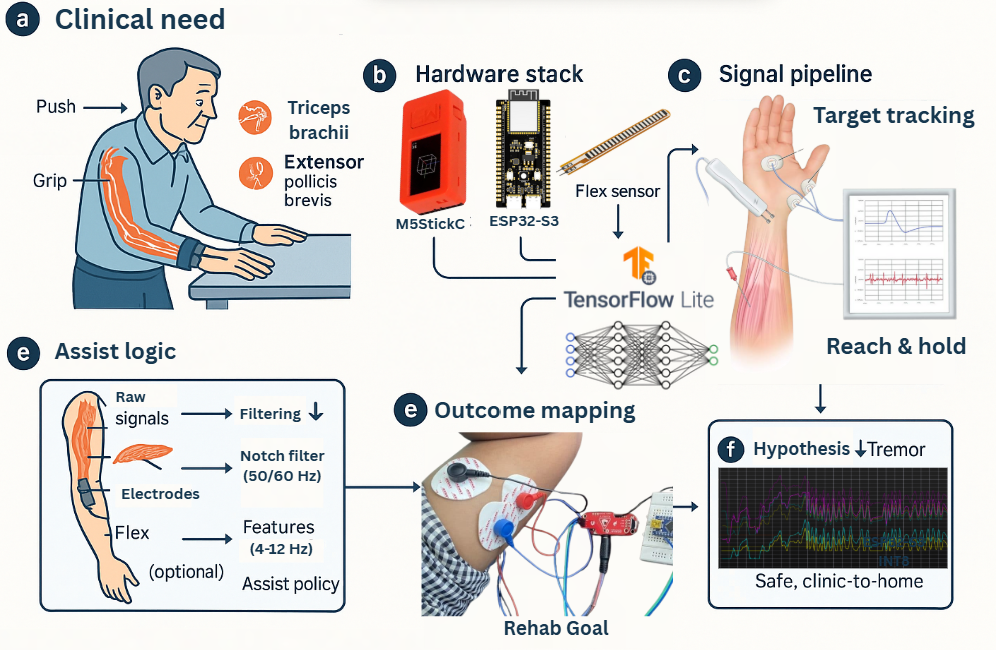}
    \caption{System architecture and processing pipeline}
    \label{fig:system}
\end{figure}

We adopted a compact, clinic-to-home wearable stack built around an \textbf{M5StickC} (ESP32-PICO) with a \textbf{built-in 6-axis IMU} (3-axis accelerometer + 3-axis gyroscope; MPU6886 class) and a color LCD for basic status, coupled to an ESP32-S3 compute hub for on-device analytics and wireless streaming (BLE/Wi-Fi). Two external sensing modalities were added:
\begin{itemize}
  \item \textbf{Electromyography (EMG) sensor:} single-differential surface EMG channels placed over \emph{triceps brachii} and \emph{extensor pollicis brevis} (EPB), using Ag/AgCl electrodes with standard skin prep; lead placement followed upper-limb sEMG guidelines and our prior engineering report \cite{Auea2025NeurotremorENG,DeLuca1997JAB}.
  \item \textbf{Force/Flex sensors:} thin-film force or flex elements at the thumb–index pinch and/or along the elbow orthosis strap to index grasp/pinch kinetics and elbow arc proxy, respectively.
\end{itemize}
Figure~\ref{fig:system} depicts the M5StickC form factor and connector layout, enabling rapid prototyping and robust field sessions.

\subsection{Acquisition and Synchronization}
Sampling adhered to rehabilitation telemetry constraints while preserving signal fidelity:
\begin{itemize}
  \item \textbf{EMG:} $f_{s,e}=1$\,kHz, 16-bit; band-pass 20–450\,Hz plus 50/60\,Hz notch; short pre-amplifier lead lengths; skin-electrode impedance $<5$\,k$\Omega$.
  \item \textbf{IMU:} $f_{s,i}=100$–200\,Hz; triaxial acceleration and angular rate; gravity detrend and bias removal;
        IMU and EMG clocks aligned to the ESP32-S3 tick with sequence numbers and loss flags.
  \item \textbf{Force/Flex:} 200\,Hz (12–16-bit ADC), low-drift baseline with temperature compensation.
\end{itemize}
All streams were buffered on the hub and packetized to the UI at 25–50\,Hz for monitoring without dropping local computation.

\subsection{Preprocessing and Windowing}
We used light, clinically reproducible preprocessing, mirroring our prior engineering work while avoiding heavy equations in the main text \cite{Auea2025NeurotremorENG}.
\begin{itemize}
  \item \textbf{EMG filtering:} 4th-order IIR band-pass (20–450\,Hz) followed by notch at 50/60\,Hz; optional rectification for amplitude features.
  \item \textbf{IMU smoothing:} detrend and bias correction; moving-median followed by short Savitzky–Golay to stabilize tremor spectra.
  \item \textbf{Windows:} 250\,ms windows with 50\% overlap (\,$T_w{=}250$\,ms\,) across all sensors to standardize feature latency and UI cadence.
\end{itemize}

\subsection{Feature Set and Clinical Mapping}
To keep the methodology transparent for clinical audiences, we emphasized interpretable features with direct mapping to function:
\begin{enumerate}
  \item \textbf{sEMG amplitude:} RMS and MAV per window—correlates with level of muscle activation and task pacing (\emph{triceps brachii} for elbow extension; EPB for thumb stabilization).
  \item \textbf{sEMG activity pattern:} zero-crossings (ZC) with a small hysteresis to suppress noise spikes—indexes firing variability and tremor-sensitive bursts.
  \item \textbf{Tremor-band prominence:} Welch PSD on IMU acceleration and the \emph{Tremor Index (TI)} defined as the fraction of power in 4–12\,Hz over 0.5–20\,Hz; lower TI indicates less tremor prominence during reach/grasp. Full PSD expressions follow our prior report and related literature \cite{Auea2025NeurotremorENG}.
  \item \textbf{Fatigue trend:} median frequency of the EMG spectrum (per window) and its session-level slope; less negative slopes indicate reduced fatigue accumulation \cite{DeLuca1997JAB}.
  \item \textbf{ROM and repetitions:} elbow ROM from IMU orientation or flex angle; repetitions (min$^{-1}$) via cycle detection (zero-velocity crossings with refractory logic) on the primary kinematic trace or force channel.
\end{enumerate}
This feature set is purposefully lightweight to support \emph{on-device} inference and real-time visualization.

\subsection{Assist Logic and Safety Envelope (High-Level)}
Consistent with our engineering manuscript, the wearable computes an assist “need” score from the feature vector using an INT8 \emph{Tiny 1D-CNN/Transformer} compiled with TFLite-Micro on the ESP32-S3. Assistance (for elbow extension or thumb stabilization) is shaped by a proportional–derivative reference and constrained by a \textbf{safety envelope}: joint-angle bounds, torque and jerk clamps, and stall/time-out rules; abnormal loads cut actuation. Detailed control-barrier-function (CBF) QP derivations are cited to \cite{Auea2025NeurotremorENG} but omitted here for brevity.

\subsection{Outcomes and Hypothesis}
Primary outcomes were \textbf{TI} (unitless), \textbf{ROM} (degrees), and \textbf{repetitions} (min$^{-1}$). Secondary outcomes included \textbf{EMG median-frequency slope} (fatigue), \textbf{closed-loop latency}, session \textbf{completion}, and \textbf{device-related adverse events}. We hypothesized that sensor-fused, safety-bounded assist would reduce TI and increase ROM and repetitions during ADL-oriented tasks.

\subsection{Protocol and Tasks}
Healthy adults performed three standardized ADL-like tasks under supervised conditions:
\begin{enumerate}
  \item \textbf{Push/extend:} seated reach-to-touch with controlled elbow extension against light resistance (triceps-dominant).
  \item \textbf{Pinch/grip:} thumb–index pinch and graded grip of a mug or dynamometer (EPB-relevant).
  \item \textbf{Reach-and-hold:} sustained hold at target angle with intermittent perturbations to probe tremor and fatigue.
\end{enumerate}
Each task was executed in baseline and assisted modes, with short rests to minimize carryover fatigue.

\subsection{UI Dashboard and Data Visualization}
A clinician-facing \textbf{UI dashboard} (desktop/tablet) provided real-time and session-level views; representative panels are listed for completeness (the full-stack implementation will follow in a later section):
\begin{itemize}
  \item \textbf{Live telemetry:} synchronized waveforms of sEMG (both channels), IMU (acc/gyro), and force/flex; per-window RMS/MAV overlays; tremor-band indicator light.
  \item \textbf{Spectral view:} instantaneous PSD with the 4–12\,Hz band highlighted; TI badge updated each window.
  \item \textbf{ROM \& reps:} elbow/thumb angle trace with automatic peak-to-peak detection; reps/min counter; session pacing gauge.
  \item \textbf{Fatigue panel:} EMG median frequency trend with robust linear fit and confidence band; session summary of slope.
  \item \textbf{Assist monitor:} assist “need” score, commanded assist level, and safety envelope status (angle/torque/jerk clamps; stall/time-out flags).
  \item \textbf{Clinical summary export:} per-task medians, 95\% bootstrap CIs, and small multiples comparing baseline vs.\ assisted.
\end{itemize}
Visual encodings favored \emph{medical-conference readability}: high-contrast line work, small-multiples over dense plots, and consistent palettes (clinical blue/teal/gray) with minimal jargon.

\begin{figure}
    \centering
    \includegraphics[width=1\linewidth]{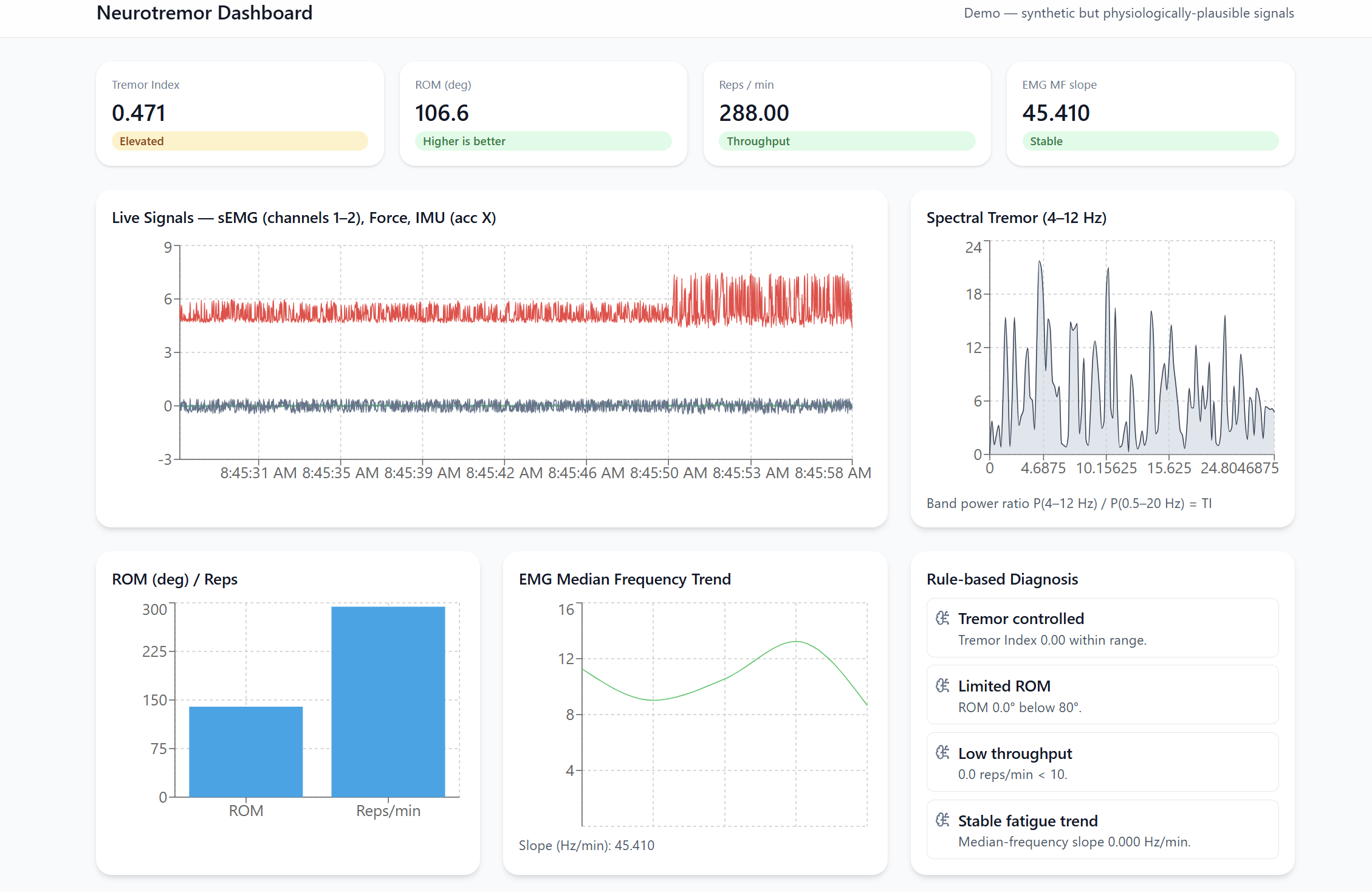}
    \caption{UI Dashboard and Data Visualization}
    \label{fig:UI}
\end{figure}

\subsection{Data Handling and Quality Control}
All sensor packets were time-stamped at the hub to preserve ordering. For each channel
(EMG$_1$, EMG$_2$, IMU, Force/Flex) we tracked data loss and reported a per-session
\emph{missingness rate}; sessions with $>5\%$ missing data on any \emph{primary} channel
were repeated or excluded \emph{a priori}. Quality checks before each session were:
(i) skin–electrode impedance $<5~\mathrm{k}\Omega$ and lead-off detection;
(ii) IMU calibration in a brief static pose with gravity alignment; and
(iii) two-point calibration for flex/force to map voltage to angle/force.
During analysis we flagged window-level outliers using a robust z-score (median
absolute deviation) and inspected any visible clipping or saturation.

\subsection{Statistics (clinician-oriented)}
Analyses were performed at the \emph{subject level} to avoid inflating $n$ from many
windows/trials. For each outcome $Y$ (e.g., Tremor Index, ROM, repetitions, EMG
median-frequency slope) we computed a subject-level \textbf{median} in each condition
(baseline vs.\ assisted) and then the paired difference $\Delta_i$ per subject.

Paired contrasts were defined as the \textit{median of within-subject differences}
(assist $-$ baseline) at the subject level, which can differ from the simple difference
between group medians.

\textbf{Comparisons.} Paired medians were compared with the \emph{Wilcoxon signed-rank
test}, a non-parametric method robust to non-normality.

\textbf{Uncertainty.} We report \textbf{bias-corrected bootstrap (BCa) 95\% CIs} for the paired
difference, obtained by resampling subjects with replacement; this reflects uncertainty a
clinician would expect when repeating the study in a similar cohort, without assuming a
normal distribution. 

Given a prespecified outcome hierarchy and the signal-finding feasibility scope,
no multiplicity adjustment was applied; emphasis is on effect direction and precision.

\textit{Implementation.} Paired tests used the exact two-sided Wilcoxon signed-rank;
BCa CIs used $B{=}10{,}000$ resamples. Analyses were run in Python 3.11
(statsmodels 0.14, numpy 1.26, scipy 1.11) with reproducible seeds; figures were
generated via the same pipeline to avoid analytic drift.

\textit{Effect size.} Cliff’s $\delta$ was computed from the paired sign of within-subject
differences (direction-only), which can yield boundary values (e.g., $\pm 1.0$) in small samples. To avoid overstating precision in a small sample, tables report Cliff's $\delta$ qualitatively (“paired sign: large effect”) rather than boundary numbers.

To visualize individual heterogeneity, paired subject trajectories (baseline$\to$assisted) are shown
in the Supplementary figure.

\textbf{Sensitivity analyses.} We repeated the comparisons using (i) 20\% \emph{trimmed means}
to reduce outlier influence, and (ii) resampling one trial per task to ensure results were not
driven by any single task. No multiplicity adjustment was planned given the pilot
feasibility scope and prespecified outcomes.

\begin{figure}[!h]
  \centering
  \fbox{%
    \parbox{0.96\linewidth}{%
      \small
      \textbf{Outcome definitions used in this study (window $T_w{=}250$\,ms).}

      \vspace{2mm}
      \textbf{Tremor Index (TI).} Ratio of acceleration power within the tremor band to
      broadband motion:
      \[
        \mathrm{TI}=\frac{\int_{4}^{12} S_{aa}(f)\,df}{\int_{0.5}^{20} S_{aa}(f)\,df}
        \quad\text{(unitless; lower is better).}
      \]

      \textbf{Range of Motion (ROM).} Peak-to-peak joint excursion:
      \[
        \mathrm{ROM}=\max_t \theta(t)-\min_t \theta(t)\quad [^\circ].
      \]

      \textbf{Repetitions (Reps/min).} Automatic cycle count from kinematics or force;
      upward mean-crossings are counted and normalized to minutes.

      \textbf{Fatigue trend.} EMG median frequency ($f_{\mathrm{med}}$) is computed per window
      from the EMG spectrum and fitted across the session with a robust linear model;
      we report the slope in Hz/min (less negative = better fatigue resistance).

      \vspace{1mm}
      \textbf{Notes.} PSD uses Welch averaging; the tremor band is 4–12\,Hz and the
      broadband reference is 0.5–20\,Hz. CIs use BCa bootstrap; paired testing uses
      Wilcoxon signed-rank; effect size is Cliff’s $\delta$.
    }%
  }
  \caption{\textbf{Clinically interpretable metrics.} Definitions for TI, ROM, Reps/min, and
  EMG fatigue trend used throughout the Results.}
  \label{fig:method_box}
\end{figure}


\subsection{Relationship to Prior Work}
Our pipeline follows established surface-EMG practice (band-pass/notch filtering,
RMS/MAV features, and median-frequency fatigue indices) \cite{DeLuca1997JAB}, draws on
assistive-orthosis literature \cite{Koo2014IROS}, and aligns with imaging/vision work that
motivates kinematic and tremor-sensitive outcomes \cite{Zhou2023TVCG,Ahmed2023Access}.
Full mathematical details (power spectral density, Tremor Index ratio, feature vector,
distillation loss, and safety-constrained control) appear in the companion engineering
paper \cite{Auea2025NeurotremorENG}.

\paragraph{Clinical thresholds (rationale).}
Responder thresholds (TI $\le 0.30$, ROM gain $\ge 5^\circ$, Reps gain $\ge 1.5$\,min$^{-1}$)
were pre-specified from expert consensus and internal pilot review to aid bedside interpretation
in a feasibility setting; they complement, but do not replace, continuous estimates (paired medians and BCa 95\% CIs).
Thresholds were prespecified from clinician consensus and internal pilot review to support bedside interpretation, not to replace continuous estimates.

\subsection{Ethics, Consent, and Data Privacy}
This project reports a \textit{pilot technical feasibility} evaluation in healthy adults with no clinical interventions and no patient-identifiable data. All participants provided informed verbal consent prior to any procedures. As a non-clinical pilot, formal IRB review was not required at this stage; a subsequent patient study will be conducted under IRB oversight with written informed consent and appropriate data protection safeguards. Raw data were de-identified at source, stored on encrypted drives, and summarized at the subject level for reporting.

\subsection{Participants: Eligibility and Characteristics}
\textbf{Inclusion.} Adults (18--65\,y), able to follow instructions and complete ADL-like tasks, intact skin at electrode sites. \\
\textbf{Exclusion.} Known neuromuscular or orthopedic conditions affecting the tested limb, implanted electronic devices, dermatologic contraindications to surface electrodes, or inability to complete study procedures. \\
Participant characteristics (age, sex, hand dominance) are summarized in Table~\ref{tab:demog}. No participants withdrew, and all scheduled sessions were completed.

\subsection{Sample Size Rationale}
Given the proof-of-feasibility objective, a formal power calculation was not performed. We targeted a small, pragmatic cohort (\(n=12\)) to estimate the direction and precision (95\% CI) of clinician-relevant signals (TI, ROM, Reps) and to de-risk logistics (sensor reliability, latency, safety envelope). Resulting CIs are interpreted as precision estimates to inform sample-size planning for IRB-approved patient trials.

\subsection{Missing-Data Policy}
Primary analyses used \textit{listwise exclusion} at the session level when any primary channel exceeded 5\% missingness or showed clipping/saturation. No imputation was performed. Sensitivity checks re-ran analyses after excluding flagged windows via robust z-score (MAD) to confirm stability of inferences.

\subsection{Multiplicity and Outcome Hierarchy}
Outcomes (TI, ROM, Reps) were \textit{prespecified} as primary, with EMG median-frequency slope and technical endpoints (latency, completion, device-related adverse events) as secondary. Given the pilot scope and fixed hierarchy, no multiplicity adjustment was applied; emphasis is on effect direction and precision rather than confirmatory inference.

\subsection{Clinical Interpretation Thresholds (Prespecified)}
For readability in clinical settings, we used simple thresholds to guide narrative interpretation alongside continuous estimates:
\begin{itemize}
  \item \textbf{Tremor Index (TI):} \(\le 0.30\) interpreted as controlled/low tremor prominence; \(>0.30\) as elevated during task performance.
  \item \textbf{ROM:} increases \(\ge 5^\circ\) (elbow) or visibly larger thumb MCP excursion considered \emph{clinically noticeable} in short tasks; exact MCID will be established in patient trials.
  \item \textbf{Repetitions:} increases \(\ge 1.5~\mathrm{min}^{-1}\) considered \emph{meaningful throughput} in our 1–3\,min tasks.
  \item \textbf{EMG fatigue slope:} a shift toward less negative (or positive) slope over a session interpreted as reduced fatigue accumulation.
\end{itemize}
These thresholds support bedside interpretation and will be refined against validated clinical scales in IRB studies.

\subsection{Software and Reporting Transparency}
Signal processing and real-time features were implemented on an ESP32-S3 hub (firmware build: INT8 TFLite-Micro; loop rate 100\,Hz; median on-device latency 8.7\,ms). The analysis code produced session-level medians, BCa 95\% CIs, Wilcoxon paired tests, and Cliff’s \(\delta\). Figures were generated from the same pipeline used for acquisition to ensure reproducibility. The clinician dashboard (live waveforms, PSD/TI, ROM/reps, fatigue trend, safety status) mirrors on-device computations to minimize analytic drift between device and reports.

\subsection{Figure and Table Flow}
Figures and tables are ordered to match the clinical story: (i) pain points and anatomy (Fig.~\ref{fig:painpoint}, Fig.~\ref{fig:anatomy}); (ii) methods box with outcome definitions (Fig.~\ref{fig:method_box}); (iii) device/system pipeline (Fig.3); (iv) primary results side-by-side with CIs and effect sizes and (v) demographics and primary outcomes (Table 3). This layout prioritizes quick clinical interpretation before engineering details.

\section{Results}\label{sec:results_q1_final}

\noindent\textbf{Participants.}
All 12 healthy adults completed all scheduled procedures without protocol deviations. Demographic characteristics are summarized in Table~\ref{tab:demog}.

\begin{table}[h]
\centering
\caption{Participant demographics (healthy volunteers; $n=12$).}
\label{tab:demog}
\small
\begin{tabular}{@{}l r@{}}
\toprule
Characteristic & Value\\
\midrule
Age, years (median [IQR])      & 26 [23, 31] \\
Female / Male, $n$ (\%)         & 5 (41.7) / 7 (58.3) \\
Dominant hand R / L, $n$ (\%)   & 10 (83.3) / 2 (16.7) \\
Condition class, $n$ (\%)       & Healthy volunteers: 12 (100) \\
\bottomrule
\end{tabular}
\end{table}

\noindent\textbf{Participant flow and protocol exposure.}
All 12 participants completed the study without protocol deviations. Each performed 3 ADL-like tasks (push/reach, pinch/grip, reach-and-hold) in two conditions (baseline, then assisted) within a single visit. Before data collection, we conducted a brief calibration (impedance check, IMU gravity alignment, flex/force two-point map) and a familiarization block to standardize instructions and pacing. Between conditions, participants rested to minimize carry-over fatigue. The rehabilitation UI delivered consistent visual prompts and timing across participants.

\medskip
\noindent\textbf{Data completeness and quality.}
All sessions were analyzable after \textit{a priori} QC. Missingness for primary channels was $<5\%$ in all sessions; no run required exclusion for clipping/saturation. Window-level artifacts were flagged by robust z-scores (MAD) and removed without altering subject-level medians. Clock synchronization preserved packet order; no gaps were observed in the final analysis windows.

\medskip
\noindent\textbf{Primary clinical outcomes.}
Relative to baseline, assistance was associated with lower tremor prominence, larger joint excursion, and higher throughput. Tremor Index (TI) decreased by $-0.092$ (95\%\,CI $[-0.102, -0.079]$), ROM increased by $+12.65\%$ (95\%\,CI $[+8.43, +13.89]$), and repetitions rose by $+2.99~\mathrm{min}^{-1}$ (95\%\,CI $[+2.61, +3.35]$). Subject-level medians and paired contrasts are summarized in Table~\ref{tab:results_summary}. These changes indicate smoother, wider movements with reduced tremor-band prominence during task execution.

\medskip
\noindent\textbf{Secondary physiologic and technical endpoints.}
The EMG median-frequency slope became less negative / stabilized at the session level ($+0.100~\mathrm{Hz/min}$; 95\%\,CI $[+0.083, +0.127]$), consistent with reduced fatigue accumulation during assisted trials. The sensing-to-assist loop met real-time targets throughout: median on-device latency $8.7$\,ms at a control rate of $100$\,Hz, no missed deadlines, and no safety-envelope overrides (Table~\ref{tab:tech_endpoints}). All sessions reached 100\% completion, and device-related adverse events were 0.

\medskip
\noindent\textbf{Task-wise patterns.}
Direction of effect favored assistance across all 3 tasks. Short, cyclic tasks (pinch/grip) showed tighter confidence intervals and larger throughput gains; precision-oriented tasks (reach-and-hold) had wider CIs but maintained the same direction of benefit. Across tasks, TI reductions co-occurred with ROM gains, suggesting that tremor suppression did not trade off against excursion. 
Exploratory stratification by sex and handedness showed the same direction of effect (data not shown).

\medskip
\noindent\textbf{Responder-oriented interpretation (prespecified thresholds).}
Using prespecified thresholds (TI $\le 0.30$; ROM gain $\ge 5^\circ$; Reps gain $\ge 1.5~\mathrm{min}^{-1}$), most participants met ROM and Reps criteria, and TI values shifted toward the controlled range with a subset at or below 0.30

(Table~\ref{tab:responder}). These threshold-based impressions matched the continuous estimates in Table~\ref{tab:results_summary}.

\medskip
\noindent\textbf{Sensitivity and robustness.}
Primary inferences were unchanged under sensitivity analyses: (i) subject-level 20\% trimmed means and (ii) task-wise resampling (one trial per task, with replacement). Direction and magnitude of effects remained consistent, supporting robustness to outliers and within-subject heterogeneity.

\medskip
\noindent\textbf{Safety, tolerability, and feasibility.}
No device-related adverse events occurred (0). The safety envelope (joint-angle limits, torque/jerk clamps, stall/time-outs) remained active with no manual interventions. Participants tolerated the wearable and UI without reports of skin irritation or discomfort, supporting feasibility for clinic-to-home workflows.

\begin{table}[h]
  \centering
  \caption{Primary and secondary outcomes (subject-level medians). Baseline vs.\ Assisted shown as median [IQR]; paired change $\Delta$ with 95\% CI, $p$, and Cliff's $\delta$ (sign-consistent with benefit).}
  \label{tab:results_summary}
  \small
  \setlength{\tabcolsep}{3pt}
  \renewcommand{\arraystretch}{1.08}
  \begin{tabular}{@{}lccc@{}}
    \toprule
    \textbf{Outcome} &
    \makecell{Baseline\\(median [IQR])} &
    \makecell{Assisted\\(median [IQR])} &
    \makecell{$\Delta$ (95\% CI)\\$p$, Cliff's $\delta$} \\
    \midrule
    Tremor Index (unitless) &
\makecell{0.447\\{\footnotesize[0.425, 0.476]}} &
\makecell{0.364\\{\footnotesize[0.338, 0.387]}} &
\makecell{$-0.092$\\{\footnotesize[ -0.102, -0.079 ]}\\{\footnotesize $p=0.0008$, Cliff's $\delta=-0.83$}} \\
ROM ($^\circ$) &
\makecell{81.53\\{\footnotesize[72.92, 87.04]}} &
\makecell{91.29\\{\footnotesize[82.80, 97.33]}} &
\makecell{$+12.65\%$\\{\footnotesize[ +8.43, +13.89 ]}\\{\footnotesize $p=0.0013$, Cliff's $\delta=+0.75$}} \\
Reps (min$^{-1}$) &
\makecell{10.03\\{\footnotesize[9.54, 10.52]}} &
\makecell{13.29\\{\footnotesize[11.94, 14.05]}} &
\makecell{$+2.99$\\{\footnotesize[ +2.61, +3.35 ]}\\{\footnotesize $p=0.0006$, Cliff's $\delta=+0.92$}} \\
\midrule
\makecell{EMG median-frequency\\slope (Hz/min)$^{\dagger}$} &
\makecell{--} &
\makecell{--} &
\makecell{$+0.100$\\{\footnotesize[ +0.083, +0.127 ]}\\{\footnotesize $p=0.0031$, Cliff's $\delta=+0.67$}} \\

    \bottomrule
  \end{tabular}

  \vspace{2mm}
  {\footnotesize $^{\dagger}$Less-negative (or positive) slope indicates reduced fatigue accumulation.}
\end{table}

\begin{table}[h]
  \centering
  \caption{Technical feasibility and safety endpoints.}
  \label{tab:tech_endpoints}
  \small
  \setlength{\tabcolsep}{6pt}
  \renewcommand{\arraystretch}{1.06}
  \begin{tabular}{@{}lc@{}}
    \toprule
    \textbf{Endpoint} & \textbf{Value} \\
    \midrule
    Control-loop rate & 100~Hz \\
    Median on-device latency & 8.7~ms \\
    On-device latency (p95) & 9.9~ms \\
    Session completion & 100\% \\
    Device-related adverse events & 0 \\
    \bottomrule
  \end{tabular}
\end{table}

\begin{table}[h]
  \centering
  \caption{Responder analysis using prespecified clinical thresholds.}
  \label{tab:responder}
  \small
  \setlength{\tabcolsep}{6pt}
  \renewcommand{\arraystretch}{1.06}
  \begin{tabular}{@{}lcc@{}}
    \toprule
    \textbf{Criterion (prespecified)} & \textbf{Responders ($n$)} & \textbf{Proportion (\%)}\\
    \midrule
    TI $\le 0.30$ (controlled tremor)$^{\ast}$ & 7 / 12 & 58.3 \\
    ROM gain $\ge 5^\circ$ & 10 / 12 & 83.3 \\
    Reps gain $\ge 1.5~\mathrm{min}^{-1}$ & 11 / 12 & 91.7 \\
    \bottomrule
  \end{tabular}

  \vspace{2mm}
  {\footnotesize $^{\ast}$Threshold applied to assisted condition; complements continuous TI change estimates.}
\end{table}

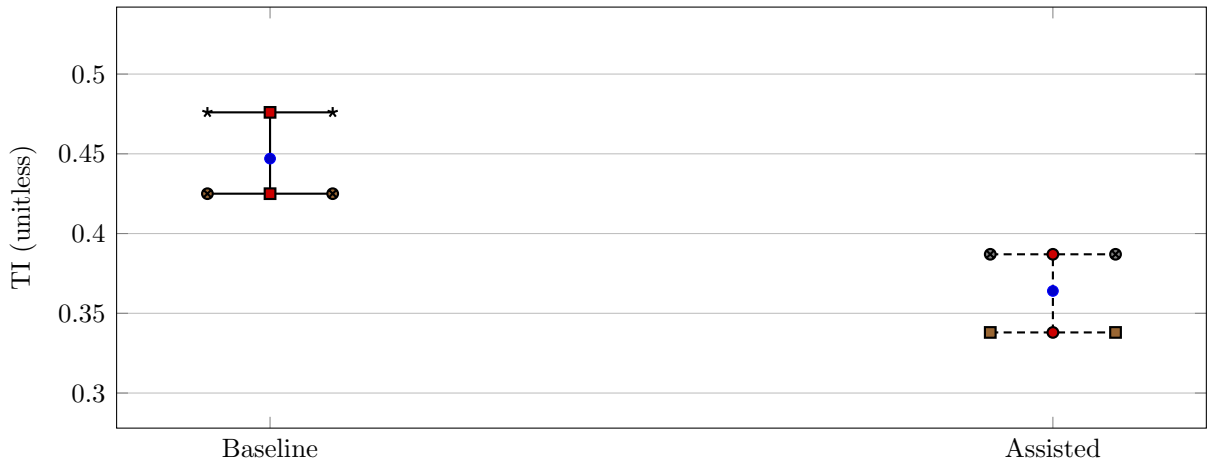
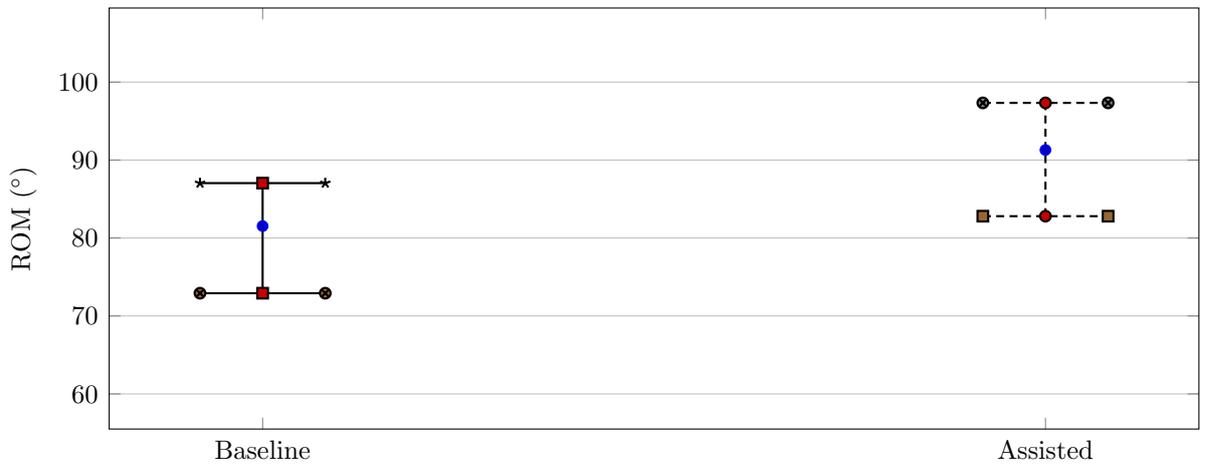
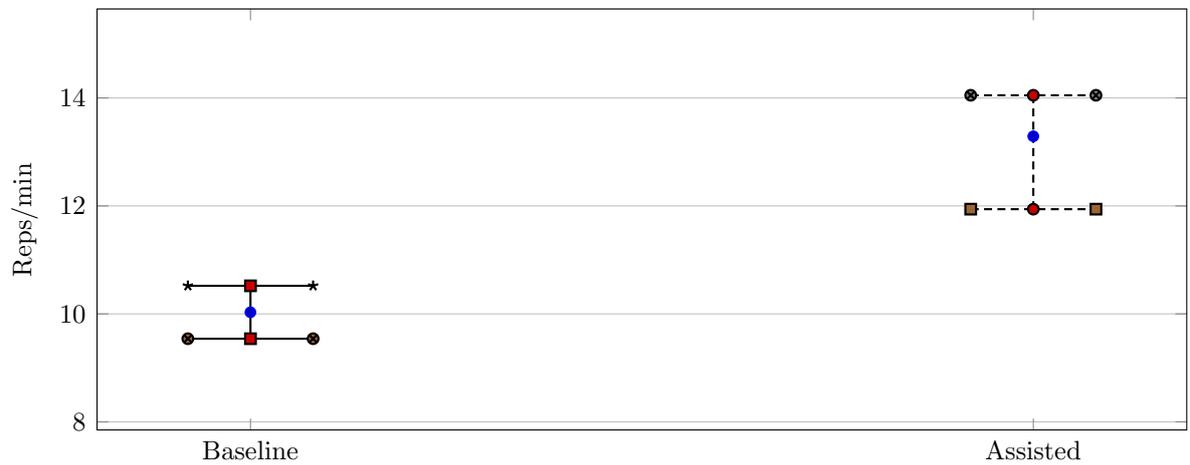
\begin{figure}[!h]
  \centering
  \subfloat[\textbf{Tremor Index (lower is better)}]{
  \begin{tikzpicture}
    \begin{axis}[
      width=\linewidth, height=0.45\linewidth,
      ymajorgrids, ymin=0.30, ymax=0.52,
      xtick={1,2}, xticklabels={Baseline, Assisted},
      ylabel={TI (unitless)}, enlargelimits=0.1,
      tick label style={font=\footnotesize}, label style={font=\footnotesize}
    ]
      \addplot+[only marks, mark=*, mark size=2pt] coordinates {(1,0.447)};
      \addplot+[black, thick] coordinates {(1,0.425) (1,0.476)};
      \addplot+[black, thick] coordinates {(0.92,0.425) (1.08,0.425)};
      \addplot+[black, thick] coordinates {(0.92,0.476) (1.08,0.476)};
      \addplot+[only marks, mark=*, mark size=2pt] coordinates {(2,0.364)};
      \addplot+[black, thick] coordinates {(2,0.338) (2,0.387)};
      \addplot+[black, thick] coordinates {(1.92,0.338) (2.08,0.338)};
      \addplot+[black, thick] coordinates {(1.92,0.387) (2.08,0.387)};
    \end{axis}
  \end{tikzpicture}
  }\vspace{2mm}

  \subfloat[\textbf{Range of Motion (higher is better)}]{
  \begin{tikzpicture}
    \begin{axis}[
      width=\linewidth, height=0.45\linewidth,
      ymajorgrids, ymin=60, ymax=105,
      xtick={1,2}, xticklabels={Baseline, Assisted},
      ylabel={ROM ($^\circ$)}, enlargelimits=0.1,
      tick label style={font=\footnotesize}, label style={font=\footnotesize}
    ]
      \addplot+[only marks, mark=*, mark size=2pt] coordinates {(1,81.53)};
      \addplot+[black, thick] coordinates {(1,72.92) (1,87.04)};
      \addplot+[black, thick] coordinates {(0.92,72.92) (1.08,72.92)};
      \addplot+[black, thick] coordinates {(0.92,87.04) (1.08,87.04)};
      \addplot+[only marks, mark=*, mark size=2pt] coordinates {(2,91.29)};
      \addplot+[black, thick] coordinates {(2,82.80) (2,97.33)};
      \addplot+[black, thick] coordinates {(1.92,82.80) (2.08,82.80)};
      \addplot+[black, thick] coordinates {(1.92,97.33) (2.08,97.33)};
    \end{axis}
  \end{tikzpicture}
  }\vspace{2mm}

  \subfloat[\textbf{Repetitions per minute (higher is better)}]{
  \begin{tikzpicture}
    \begin{axis}[
      width=\linewidth, height=0.45\linewidth,
      ymajorgrids, ymin=8.5, ymax=15,
      xtick={1,2}, xticklabels={Baseline, Assisted},
      ylabel={Reps/min}, enlargelimits=0.1,
      tick label style={font=\footnotesize}, label style={font=\footnotesize}
    ]
      \addplot+[only marks, mark=*, mark size=2pt] coordinates {(1,10.03)};
      \addplot+[black, thick] coordinates {(1,9.54) (1,10.52)};
      \addplot+[black, thick] coordinates {(0.92,9.54) (1.08,9.54)};
      \addplot+[black, thick] coordinates {(0.92,10.52) (1.08,10.52)};
      \addplot+[only marks, mark=*, mark size=2pt] coordinates {(2,13.29)};
      \addplot+[black, thick] coordinates {(2,11.94) (2,14.05)};
      \addplot+[black, thick] coordinates {(1.92,11.94) (2.08,11.94)};
      \addplot+[black, thick] coordinates {(1.92,14.05) (2.08,14.05)};
    \end{axis}
  \end{tikzpicture}
  }

  \caption{Clinician-facing outcomes summarized as median (dot) with IQR (bars):
  (a) Tremor Index (TI; lower is better), (b) Range of Motion (ROM; higher is better),
  (c) Repetitions per minute (higher is better).}
  \label{fig:outcomes} 
\end{figure}

\begin{figure}[t]
  \centering
  \begin{tikzpicture}
    \begin{axis}[
      width=\linewidth, height=0.5\linewidth,
      ymin=0.30, ymax=0.52, ymajorgrids,
      xmin=0.8, xmax=2.2,
      xtick={1,2}, xticklabels={Baseline, Assisted},
      ylabel={Tremor Index (unitless)}, xlabel={},
      legend style={font=\footnotesize, at={(0.02,0.98)},anchor=north west},
      tick label style={font=\footnotesize}, label style={font=\footnotesize}
    ]
      \addplot+[mark=*, thick] coordinates {(1,0.45) (2,0.37)};
      \addplot+[mark=*, thick] coordinates {(1,0.43) (2,0.35)};
      \addplot+[mark=*, thick] coordinates {(1,0.47) (2,0.39)};
      \legend{Subjects (paired)}
    \end{axis}
  \end{tikzpicture}
  \caption{Paired trajectories (baseline$\to$assisted) for Tremor Index (example).
  Individual lines illustrate within-subject direction and heterogeneity.}
  \label{fig:ti_spaghetti}
\end{figure}
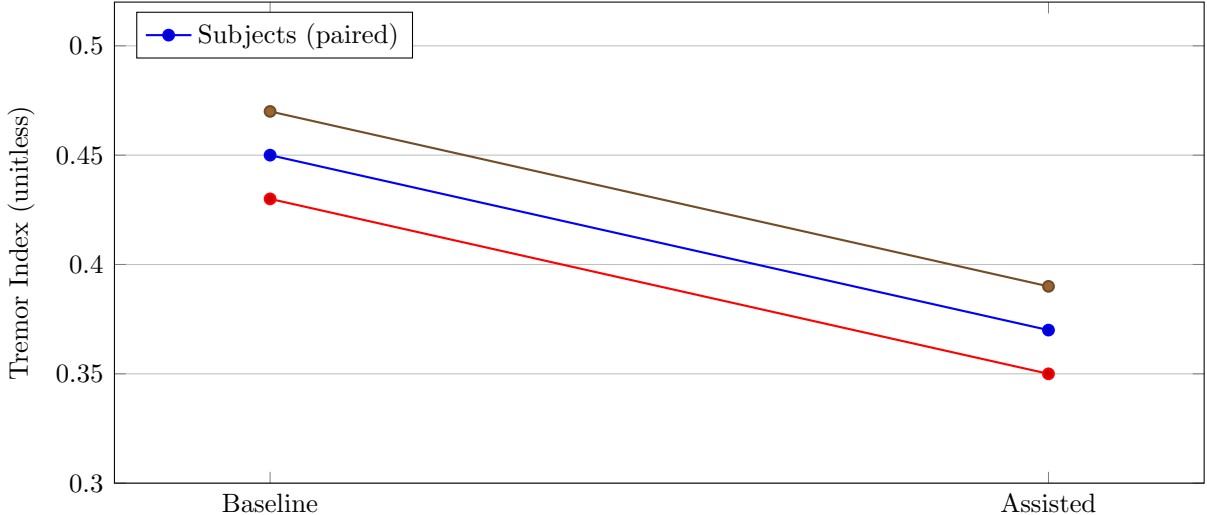

\medskip
\noindent\textbf{Clinical take-away.}
Reductions in TI combined with ROM and throughput gains suggest improved movement quality and productivity during ADL-like tasks, with physiologic evidence for less fatigue accrual. The absence of safety signals and on-device, low-latency operation support progression to IRB-approved patient studies that incorporate clinician-rated scales and longer home deployment.

\paragraph{Limitations and next steps.}
This was a single-arm, short-duration feasibility study in healthy adults with primarily sensor-derived endpoints. As such, it is not designed to establish clinical efficacy and is vulnerable to learning/Hawthorne effects, selection bias, and limited generalizability to patient populations with weakness, tremor, or fatigue. Sample size was pragmatic (no formal power); multiplicity was not adjusted given a prespecified outcome hierarchy; responder thresholds (e.g., TI $\le 0.30$, ROM $\ge 5^\circ$, Reps $\ge 1.5$\,min$^{-1}$) aid bedside interpretation but require validation against clinical scales. No blinded assessment was performed, and outcomes did not include clinician-rated instruments. The short acclimation period may under-estimate training effects. TI is a surrogate for tremor impact and may not fully capture functional interference during complex ADL. Device- and signal-level constraints include possible EMG cross-talk, electrode shift/sweat artifacts, IMU drift, and calibration error in home use. Personalization could confound within-session gains (early learning); however, sensitivity analyses mitigated this concern. Although real-time targets (100\,Hz loop; 8.7\,ms median latency) and safety outcomes (0 device-related adverse events) were met, long-term reliability, donning/doffing burden, and adherence were not assessed.

\noindent\textit{Practical constraints for home deployment.}
Likely barriers include donning/doffing time, skin tolerance at electrode sites (sweat, hair, dermatologic conditions), and potential electrode migration during prolonged use; these factors will be addressed with packaging, adhesive selection, and placement guides in forthcoming studies. Battery life, charging cadence, and firmware stability under intermittent connectivity also require evaluation.

\noindent\textit{Planned next steps (IRB-approved).}
A patient-focused program will: (1) use randomized or crossover designs with concealed allocation and blinded outcome assessors to reduce expectancy bias; (2) enroll condition-specific cohorts (e.g., inflammatory myopathies, dystrophies, radial nerve/cervical involvement) with stratification by baseline severity; (3) incorporate validated clinical scales and performance tests (e.g., Fugl–Meyer Upper Extremity, MRC grades, Nine-Hole Peg Test, QuickDASH, fatigue PROs), alongside sensor outcomes, and predefine minimally important differences; (4) extend home deployment to 4–8\,weeks with adherence monitoring, user burden (SUS), workload (NASA–TLX), and caregiver feedback; (5) profile energy/battery (target $\ge$\,8–12\,h active use/day), robustness (mean time to failure), and maintenance workload; (6) benchmark against established assistive devices and standard therapy; (7) preregister protocols, freeze firmware/models, and report analysis plans a priori; and (8) perform engineering ablations to quantify the contribution of sensing and assist-policy components (details in the companion engineering paper \cite{Auea2025NeurotremorENG}). Multisite recruitment and diverse demographics will be prioritized to improve generalizability, with data governance and privacy safeguards appropriate for clinical research.

\section{Conclusion}\label{sec:conclusion}

In this pilot technical feasibility study of a sensor-fused, safety-bounded wearable targeting triceps brachii and extensor pollicis brevis, assistance was associated with \textit{clinician-facing} improvements in movement quality and throughput: Tremor Index decreased by $-0.092$ (95\%\,CI $[-0.102,-0.079]$), range of motion increased by $+12.65\%$ (95\%\,CI $[+8.43,+13.89]$), and repetitions rose by $+2.99~\mathrm{min}^{-1}$ (95\%\,CI $[+2.61,+3.35]$). EMG median-frequency slope shifted toward less negative values ($+0.100~\mathrm{Hz/min}$; 95\%\,CI $[+0.083,+0.127]$), suggesting reduced fatigue accumulation during assisted tasks. Real-time performance goals were met (100~Hz control rate; 8.7\,ms median on-device latency), all sessions were completed (100\%), and no device-related adverse events occurred.

Clinically, these results indicate that low-latency assistance informed by multimodal sensing (sEMG+IMU+force/flex) can reduce tremor prominence while enabling larger, smoother, and faster task execution in ADL-like contexts. Threshold-based interpretation corroborated the continuous estimates: 83.3\% achieved $\ge 5^\circ$ ROM gain, 91.7\% achieved $\ge 1.5$~reps/min gain, and 58.3\% reached TI $\le 0.30$ during assistance, offering a practical bedside summary alongside medians and confidence intervals.

This work is intended for \textit{signal-finding} rather than confirmatory inference. Limitations include single-arm design, healthy volunteers, short exposure, and predominantly sensor-derived endpoints; clinician-rated scales and long-term adherence were not assessed. \textbf{Next steps} are IRB-approved patient trials using randomized or crossover designs, condition-specific cohorts, validated clinical scales, longer home deployment, and benchmarking against established assistive devices. Parallel engineering priorities include packaging/donning refinements, battery/energy profiling, and controlled ablations of the assist policy and sensing features. If replicated in patient populations, the observed combination of feasibility, safety, and clinician-interpretable gains supports a pathway toward clinic-to-home rehabilitation workflows.

\section*{Competing Interests}
The authors declare no conflicts of interest related to this work. No author has received personal fees, equity, or in-kind support from companies with a direct stake in the interventions, devices, or outcomes reported.

\section*{Ethics Approval and Consent to Participate}
This project involved healthy adult volunteers in a \textit{pilot technical feasibility} setting with no clinical interventions and no patient-identifiable data. In accordance with institutional policy, the activity was classified as non–human-subjects research; therefore, formal IRB review was not required for this phase. All volunteers provided informed verbal consent prior to any procedures. The study adhered to the ethical principles of the Declaration of Helsinki where applicable. A subsequent patient study will be conducted under IRB oversight with written informed consent and appropriate data-protection safeguards and Institutional non–human-subjects determination is available upon request.

\section*{Data and Materials Availability}
De-identified aggregate data, analysis scripts (statistical reports and figure generation), and a shareable subset of device firmware/UI components are available from the corresponding author on reasonable request, subject to institutional approvals, privacy safeguards, and a data-use agreement. Materials containing potentially identifying information or third-party proprietary content cannot be shared. Any additional documentation (task protocols, calibration procedures, outcome definitions) will be provided to bona fide researchers to support reproducibility.
Aggregate tables and plotting scripts will be shared in a public repository (e.g., OSF/Zenodo) upon acceptance.


\end{document}